\documentclass[usenatbib,twocolumn]{aastex701}
\usepackage{natbib}
\shorttitle{The Lightcurve of V854 Cen}
\shortauthors{Clayton et al.}

\begin{document}

\title{The Historical Lightcurve of the R Coronae Borealis Star, V854 Cen, from 1890 to 2026}



\author[0000-0002-0141-7436]{Geoffrey C. Clayton}
\affiliation{Space Science Institute,
4765 Walnut St., Suite B
Boulder, CO 80301, USA}
\email[show]{
gclayton@spacescience.org}

\begin{abstract}
The R Coronae Borealis (RCB) Star, V854 Cen, was not discovered until the 1980's even though it is 7th magnitude at maximum light. This is because it was in a faint state due to the presence of thick circumstellar dust clouds for nearly a century. RCB stars are known for having deep declines of up to 9 magnitudes at irregular intervals. 
The declines are caused by dust clouds which block the light from the star.
The historical lightcurve of V854 Cen before discovery was investigated by examining plates taken by Harvard College Observatory and other observatories beginning in 1889. These observations show that V854 Cen was well below 7th magnitude from 1890 to 1980, often as faint as 16th or 17th magnitude.

\end{abstract}


\section{Introduction}
V854 Cen is the third brightest R Coronae Borealis (RCB) star (V=7.1, B-V=0.5), but it was not discovered as a variable star until 1986 \citep{1986IAUC.4233....3M}. 
It is not in the HD or SAO catalogs because the star was quite faint when those catalogs were produced. 

V854 Cen (CD -39 9021, NSV 6708) was first cataloged between 1885 and 1891 in the Cordoba Durchmusterung with a visual magnitude of 9.7 \citep{1894cdbp.book.....T}. Its variability was first noted on sky patrol plates of the Bamberg Southern Station \citep{1964IBVS...74....1S}. They noted a photographic magnitude of 9.7 and a variation of 0.8 mag. 

The lightcurve, spectra, and IR excess of V854 Cen all suggest that it is an RCB star \citep{1989MNRAS.240..689L,1989MNRAS.238P...1K}. RCB stars are a rare class of variables having about 150 members \citep{2025MNRAS.537.2635C}. They are hydrogen deficient, carbon supergiants, which are thought to result from a binary, white-dwarf merger event \citep{2012JAVSO..40..539C}. The RCB stars are characterized by large declines in brightness at irregular intervals caused by clouds of dust around the star. 

When V854 Cen was ``discovered", some of the 
Harvard College Observatory (HCO) plates were examined. The plates from 1913-1921 and 1933-1952 show V854 Cen to be B$\sim$10 mag in mid-1917 and visible on some plates at B=11-13 mag during 1919-1937. But on most plates the star was $\gtrsim$13 mag and not present \citep{1986IAUC.4245....2M}. 
\citet{1986IAUC.4241....2W} reported that plates from 1935-1936 and 1952-1953 show little variation and a mean mpg = 11.5. But they may have been confused by a nearby star, UCAC2 14839501 (V=11.0, B=12.0) which lies 33\arcsec~to the south of V854 Cen.

\begin{figure*}
\centering
\includegraphics[width=6in]{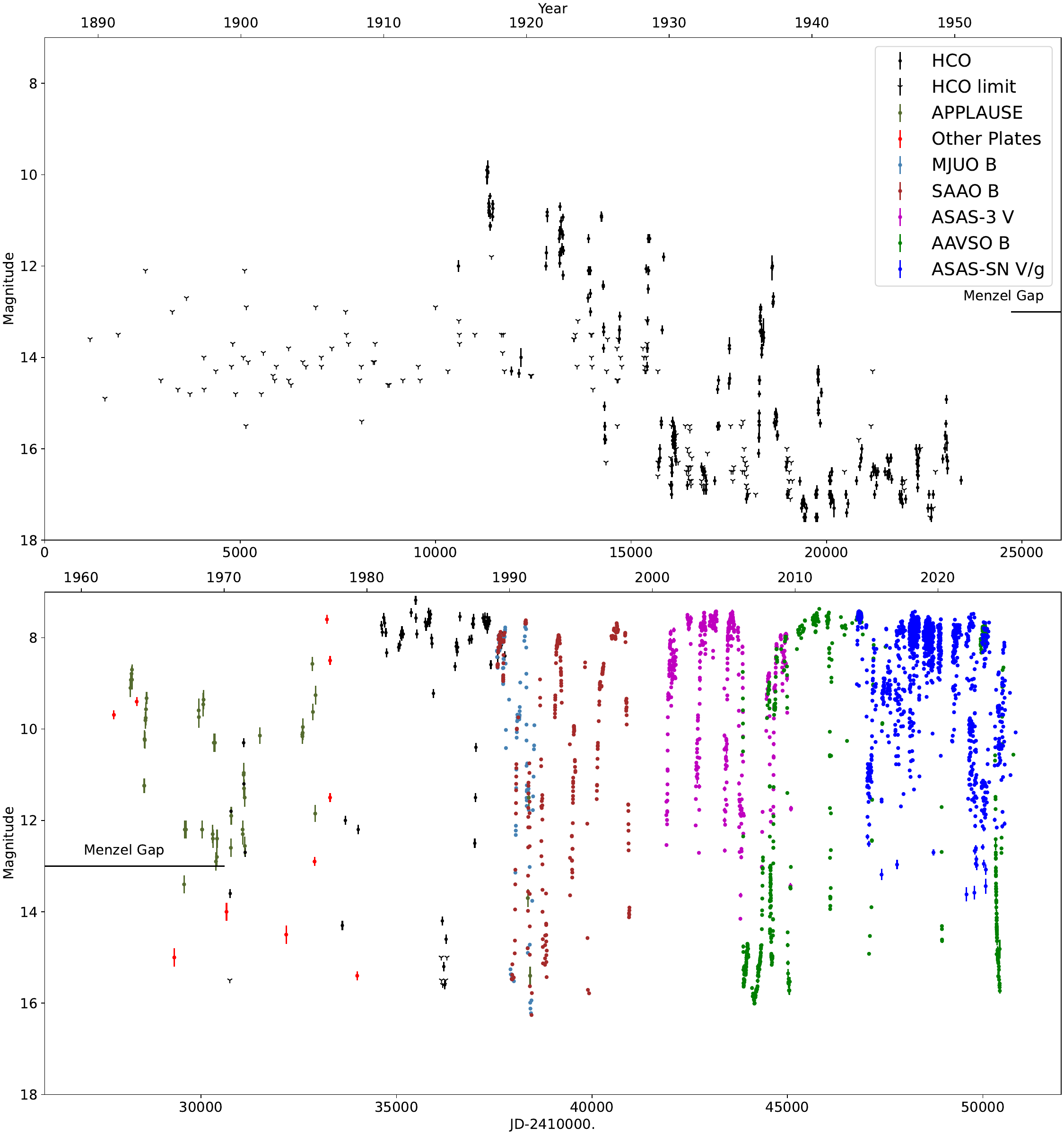}
\caption{The lightcurve of V854 Cen from 1890 to 2026. For the sources of the photometry, see the text.}
\label{fig:lightcurve}
\end{figure*}

\section{Data}
The Digital Access to a Sky Century @ Harvard (DASCH) project scanned and performed photometry on all the glass plates at HCO \citep{2012IAUS..285...29G,2025arXiv250112977W}. 

V854 Cen is in the field of 2634 HCO plates. The DASCH lightcurve and cutouts of the field around the star were produced using the daschlab (DR7) modules. The first HCO plate was obtained in 1889 and the last in 1989. No plates were obtained during the Menzel gap from 1954 to 1970 \citep{2012IAUS..285..243G}. 
The lightcurve of V854 Cen produced using the DASCH photometry shows that it was well below its present maximum brightness until about 1980. But a visual inspection of the plates, revealed serious problems.

Many of the observations are clustered around 12th magnitude. The close companion to V854 Cen, UCAC2 14839501, was often measured by mistake when V854 Cen was faint. When the seeing is good and V854 Cen is fainter than about 12th magnitude, the DASCH photometry is usually correct. 
But when V854 Cen is brighter than this, the two stars begin to merge on the plates.
On many plates, the stars are trailed, have poor seeing, or look like birds. Good photometry can still be done on some of these plates by eye.
On many of the plates, the WCS is wrong and the companion star was wholly or partially measured instead of V854 Cen. 
At maximum, V854 Cen is saturated on the plates and the photometry is not accurate. The DASCH photometry hits a ceiling at 8th magnitude. In reality, the star can be as bright as 7th magnitude. 

To correct these issues, each plate was inspected visually.
A photometric sequence for the V854 Cen field was produced using B magnitudes from the APASS secondary standards \citep{2012JAVSO..40..430H}. 
The color response of the photographic plates is very similar to the Johnson B filter \citep{2013PASP..125..857T}.
 
On plates with good seeing and tracking where V854 Cen and UCAC2 14839501 are well separated and the WCS is correct, the DASCH photometry was adopted.
On plates where V854 Cen is much brighter than UCAC2 14839501 and the two stars are merged and saturated, Photutils was used to calculate the FWHM of the images of V854 and stars in the photometric sequence to get good photometry \citep{1981PASP...93..253S}.
For the remaining plates, the brightness of V854 Cen was estimated by eye by comparing it to other stars in the photometric sequence which covered B magnitudes from 7 to 17. 

\section{Results}
The final lightcurve is shown in Figure~\ref{fig:lightcurve}. 
The sources of the photometry include the HCO plates, additional plates available from the Archives of Photographic PLates for Astronomical USE (APPLAUSE)\footnote{https://www.plate-archive.org}, photometry from Mount John University Observatory (MJUO)\citep{1989MNRAS.240..689L}, the South African Astronomical Observatory (SAAO)\citep{1999AJ....117.3007L},
the AAVSO, ASAS-3 \citep{1997AcA....47..467P}, ASAS-SN \citep{2014ApJ...788...48S,2023arXiv230403791H}, other random plates including sky survey plates listed in \citet{1986IBVS.2928....1M}, and the 1st and 2nd Cape Photographic Catalogs, and 
USNO-A1.0. The ASAS-3 V photometry was corrected to B by adding the B-V =0.5 for V854 Cen. The ASAS-SN V and g photometry was not corrected since it is extremely saturated. 

This lightcurve is very different from the one produced by the DASCH photometry. 
V854 Cen does not appear on any plate until 1915. It is typically 14th magnitude or fainter from 1890 to 1915. Around 1917, V854 Cen brightens to 10th magnitude briefly before slowly sinking to 16-17th magnitude where it remained from 1930 to 1950 with only one brief brightening. It began to slowly brighten toward maximum sometime during the Menzel gap. Other plates taken at southern stations of European observatories fill in some of this period. V854 Cen returned to its maximum brightness, B$\sim$7, for the first time around 1980. Since 1980, V854 Cen has behaved as normal, active RCB star with frequent declines followed by a return to maximum light. The lightcurve behavior of V854 Cen is unique among RCB stars.

The HCO photometry and plate limits are available at
https://doi.org/10.5281/zenodo.18992573. The photometry of the APPLAUSE plates is available at https://doi.org/10.5281/zenodo.18992848.\\



\bibliography{everything2}
\end{document}